# A Blockchain Consensus Mechanism for Distributed Electricity Trading

*Shanglin Yang, Guo Chen*, Member, IEEE, Shiping Chen, Member, IEEE*

***Abstract*—Distributed power systems complement centralized grids by coordinating distributed energy resources (DERs) to achieve regional energy self-sufficiency. Scaling such systems raises four persistent challenges: decentralized coordination, fair economic settlement, trustworthy operation, and system optimization, all without a central authority. This paper proposes Proof of Energy (PoE), a blockchain consensus mechanism that addresses these challenges through cryptographically secured, contribution-proportional node selection. In PoE, block generation rights are tied directly to real-world energy contributions, enabling distributed consensus without centralized dispatch. An Energy Contribution Unit (ECU) model is introduced to map heterogeneous energy services onto a unified value metric via scarcity-weighted normalization. A Verifiable Random Function (VRF)-based proposal mechanism then ensures selection probability is strictly proportional to node contribution, preserving fairness and resisting manipulation. Case studies validate PoE across three dimensions: grid coordination, incentive fairness, and optimization efficiency. The result is a cryptographically secured, incentive-compatible framework for decentralized value distribution in energy systems.**

## I. INTRODUCTION

CONVENTIONAL power system architectures rely on centralized scheduling and settlement. A central control center coordinates generation, transmission, distribution, and electricity sales [1]. This model has been extensively validated in mature electricity markets. Large-scale regional markets—such as AEMO's National Energy Market (NEM) in Australia and the PJM Interconnection in the United States—demonstrate the proven effectiveness of centralized dispatch in bulk power systems [2]. However, with the large-scale integration of distributed energy resources (DERs), traditional power systems face mounting operational pressure. The aim here is not to argue for wholesale decentralization. Rather, the goal is to clarify the conditions under which each architecture holds a comparative advantage.

The first limitation is structural vulnerability caused by reliance on a central control center. Panteli et al. [3] identified the control center as a single point of failure (SPOF). Any malfunction carries the risk of paralyzing the entire system. This vulnerability is particularly acute under extreme conditions. During the Texas power crisis of February 2021, communication failures at the grid dispatch center extended both the scope and duration of the outage, leaving over 4.5 million customers without power [4]. In armed conflict, critical power infrastructure is a primary target. Destruction of central facilities can trigger grid-wide collapse [5], [6].

Centralized architecture also imposes barriers in infrastructure-constrained regions. In large parts of sub-Saharan Africa, grid infrastructure remains sparse and fragmented. According to the International Energy Agency, a substantial share of the region's population lacks reliable electricity access [7]. Under such conditions, building and maintaining a centralized dispatch center is neither technically practical nor economically justified. Distributed microgrids offer a more viable path to electrification in these contexts [8].

The second limitation is the scalability ceiling of centralized dispatch under high DER penetration. The number of active nodes in distribution systems has grown rapidly with DER deployment [9]. Computational load at the central controller scales exponentially with node count. In P2P market clearing, the matching problem reaches NP-hard complexity [10]. Both processing capacity and communication bandwidth at the dispatch center become binding constraints. Gungor et al. [11] showed that under high DER penetration, the central scheduler must resolve large-scale, high-frequency decision problems that strain the existing architecture beyond practical limits.

The transition from centralized to decentralized architecture introduces four fundamental challenges. (1) The Coordination Problem. Without a central dispatcher, distributed nodes must reach consensus on system state and coordinate actions in real time to maintain grid stability [12]. In traditional power systems, the control center issues dispatch commands directly. No equivalent coordination mechanism exists in a decentralized architecture. (2) The Economic Problem. Without a central settlement authority, the system must guarantee fair compensation across all service types [13]. Current distributed trading models cover energy generation but largely exclude ancillary services—frequency regulation, voltage support, and reserve capacity—that are essential to system stability. (3) The Trust Problem. Without a trusted third party, the system must prevent data falsification and opportunistic reporting in self-declared energy contributions and service provision [14]. Reliance on self-reported data creates incentives for strategic misrepresentation. (4) The Optimization Problem. Without a Shanglin Yang is with the School of Electrical Engineering and Telecommunications, University of New South Wales, NSW 2052, Australia and School of Energy and Materials, Shihezi University, 832000 China (e-mail: yangshanglin_he@ 163.com).

Guo Chen is with School of Electrical Engineering and Telecommunications, University of New South Wales, NSW 2052, Australia (e-mail: guo.chen@unsw.edu.au).

Shiping Chen is with Data 61 CSIRO, Australia and School of Computer Science and Engineering, University of New South Wales, NSW 2052, Australia (shiping.chen@data61.csiro.au).

central planner, the system must align individual node decisions with network-level performance objectives under fully distributed decision rules [15]. Addressing these four challenges requires a technical framework grounded in the physical and economic properties of distributed energy systems.

Blockchain is a distributed ledger that records and verifies transactions through cryptographic hash chains and consensus mechanisms, eliminating the need for a trusted intermediary [16]. Its core properties of decentralization, immutability, and traceability provide a natural trust framework for multi-party environments. Cryptographic guarantees enable trustless verification, and smart contracts execute transactions autonomously. These properties align closely with the operational requirements of decentralized microgrid trading.

Researchers have proposed numerous blockchain-based P2P trading architectures that allow microgrid prosumers to trade directly without a central control center. Tushar et al. [17] identified key benefits of blockchain-enabled P2P energy trading, including lower intermediary costs, greater transaction transparency, and improved user autonomy. Khorasany et al. [18] developed a matching-algorithm-based P2P mechanism that pairs buyers and sellers directly, improving settlement efficiency and supporting the prosumer model. However, P2P trading does not resolve the four fundamental challenges of decentralized systems identified above. Wang et al. [19] showed that, without an appropriate coordination mechanism, individually rational decisions conflict with system-level optimality. When each user maximizes personal benefit, the system may converge to a suboptimal operating state. A user may select the cheapest supplier irrespective of electrical distance, increasing transmission losses and preventing network-wide power flow optimization [20]. This tension directly instantiates the fourth optimization problem of decentralized power systems described above. Mollah et al. [21] further highlighted the trust problem: without reliable third-party oversight, enforcing honest contract compliance remains an open challenge. A separate structural gap also persists. Existing blockchain trading mechanisms focus almost exclusively on energy (kWh) transactions. Stable microgrid operation requires ancillary services including frequency regulation, voltage control, and reserve capacity, in addition to generation. Most mechanisms provide no valuation framework for these services [22]. Nodes therefore lack incentives to deliver full system support, weakening grid robustness. The conflicts inherent in existing blockchain-based decentralized trading, including coordination failure, incentive mismatch, trust deficiency, and optimization misalignment, directly correspond to the four systemic problems of decentralized power systems identified above. We argue that these problems stem from the continued application of centralized paradigms to decentralized systems. Addressing them requires a purpose-built consensus mechanism that reconstructs the coordination and incentive logic of the decentralized microgrid from first principles.

A consensus mechanism defines the rules and algorithms by which nodes in a distributed system reach agreement on system state. It plays a decisive role in determining the security, efficiency, and scalability of a blockchain system [23]. A well-designed consensus mechanism therefore offers a principled path to resolving the systemic problems of the decentralized power grid. Applying centralized design logic to a decentralized grid does not eliminate these problems; it merely reintroduces them in a new form.

Traditional consensus mechanisms, including Proof of Work (PoW) and Proof of Stake (PoS), have proven successful in the cryptocurrency domain but were not designed for energy systems and cannot adequately address the specific requirements of microgrid power trading. Proof of Work (PoW), introduced by Nakamoto [24], is the earliest and most widely adopted blockchain consensus mechanism. Nakamoto's design achieves decentralized consensus through competition based on computational power, establishing the foundation for blockchain technology. Under the PoW framework, nodes (miners) compete for block proposal rights and block rewards by solving computationally intensive but practically valueless hash puzzles. However, PoW suffers from extremely high energy consumption. Vranken [25] reported that Bitcoin mining consumes an amount of electricity annually equivalent to the total consumption of several medium-sized countries. This represents a fundamental mismatch: PoW functions as an energy black hole rather than an energy management tool, and its design orientation is directly opposed to the functional purpose of an energy system.

To address the high energy consumption of PoW, researchers proposed the PoS mechanism, in which a node's probability of being selected as a block producer is proportional to its stake in the system, typically measured by the number of tokens held. Saleh [26] provided a detailed analysis of PoS, highlighting its high energy efficiency and strong scalability. Based on the original PoS, several variants have been developed. Delegated Proof of Stake (DPoS) selects representatives to produce blocks, further improving efficiency, while Practical Byzantine Fault Tolerance (PBFT) enables faster transaction confirmation. However, PoS and its variants ground selection rights in token holdings rather than physical energy contributions. This disconnect from the actual behavior of energy system participants means that PoS-based mechanisms cannot address the physical characteristics of energy systems [27].

To address the specific requirements of energy systems, researchers have explored consensus mechanisms better aligned with energy applications. Cao et al. [28] proposed a carbon credit-based consensus mechanism that uses voluntary carbon emission reductions as the basis for block proposals. Chen et al. [29] designed a Proof of Solution consensus mechanism that replaces the meaningless hash computation of PoW with energy dispatch and trading calculation. Liu et al. [30] introduced a reputation-based consensus mechanism for energy trading, using nodes' historical transaction behavior as the evaluation metric. These studies provide valuable efforts to integrate consensus mechanisms with the characteristics of energy systems. However, most mechanisms consider only a single dimension of energy contribution and cannot provide a comprehensive assessment of node value. Many approaches cannot adapt to the real-time operational demands of energy systems, limiting their practical applicability. To systematically address the problems of decentralized power grids described above, this paper proposes Proof of Energy (PoE), a blockchain consensus mechanism that grounds consensus in actual energy contributions rather than meaningless computational competition. The core idea of PoE is to redirect the energy expenditure of conventional blockchain consensus toward productive energy generation, transforming the blockchain from an energy black hole into an energy white hole. The main contributions of this paper are as follows:

1) A block-generation-based consensus mechanism named Proof of Energy (PoE) is proposed to address the coordination problem. Nodes participate in block generation based on actual energy contributions. Distributed consensus is achieved through PoE, enabling system coordination without centralized scheduling.

2) The Energy Contribution Unit (ECU) model is designed as a unified service evaluation framework to address the incentive problem. ECU standardizes the valuation of heterogeneous energy services, including power generation, frequency regulation, voltage control, and reserve capacity. This establishes a fair value quantification system that incentivizes comprehensive system support.

3) A cryptographically secured block proposal mechanism is developed to ensure fairness and resist manipulation. Through Verifiable Random Functions (VRF), the system implements contribution-based sortition. Block proposal rights are assigned with probabilities strictly proportional to node contributions, while the inherent randomness of VRF enhances system security.

4) Case studies are conducted based on the IEEE 123-node distribution system across three representative operating scenarios to validate the mechanism's effectiveness and performance.

The remainder of this paper is organized as follows. Section II presents the PoE mechanism design. Section III presents the case studies and discusses the results. Section IV concludes the paper.

## II. Framework of the PoE-based Consensus Mechanism

The architecture and workflow of PoE are illustrated in Fig. 1. The framework operates across three logical tiers.

At the physical resource layer, heterogeneous microgrid nodes broadcast real-time operational data, including active power P, reactive power Q, voltage V, frequency f, and service status, to the distributed network. This ensures network-wide data tracking and immutability. Node types include power generation nodes such as photovoltaic, wind, and thermal units, consumer nodes, and ancillary service nodes.

The core of the framework is the PoE consensus layer. The system first computes scarcity indices for each service type. These indices dynamically calibrate the conversion factors used by the Energy Contribution Unit (ECU) model, which maps heterogeneous physical services, including power generation, frequency regulation, and voltage support, into a unified contribution value. The VRF mechanism then performs contribution-based probabilistic sortition, ensuring that each node's probability of obtaining block proposal rights is strictly proportional to its ECU contribution, thereby preventing centralized consensus.

Atop these two layers sits the blockchain and incentive layer. Smart contracts automatically execute reward allocation according to on-chain consensus results. This layer enforces the principle of contribution-proportional probabilistic reward: each node's expected reward is directly proportional to its actual system contribution. In this framework, rules are enforced by code rather than by any central authority. Long-term fairness is maintained through cryptographic verification and smart contract execution alone, without reliance on any intermediary. This eliminates intermediary rent-seeking and establishes a fair, transparent, and verifiable value distribution framework for decentralized energy systems.

### *A. Workload Measurement*

The PoE-based consensus' workload measurement informs how to aggregate quantitative measurements from diverse services in a uniform manner. To achieve a standardized metric, ECU is introduced:

$$ECU_{i,k}^{t} = W_{i,k}^{t} \times C_{k}^{t} \times Q_{i,k}^{t} \tag{1}$$

where $C_k^t$ represents the conversion factor for service k at time t, and $Q_{i,k}^t$ represents the quality factor of service k provided by node i. The conversion factor maps workloads expressed in different physical units into a unified ECU, while the quality factor captures the service's performance level.

The conversion factor is dynamically adjusted based on the scarcity conditions of the service in the system:

$$C_k^t = C_k^{base} \times (1 + S_k^t) \tag{2}$$

where $C_k^{base}$ represents the baseline conversion factor for service k, and $S_k^t$ represents the scarcity index of service k at time t in the system. The scarcity index of service is calculated by the supply and demand situation:

$$S_k^t = f_k(Sys^t, Sys^{ref}, Sys^{hist}) \tag{3}$$

Where $Sys^t$ is the operational status of the microgrid at time t. The vector contains the instantaneous frequency, voltage, load and other relevant data; $Sys^{ref}$ is the state vector for the reference system. It shows the design target values of the microgrid. It serves to evaluate the deviation from a normal operating state; $Sys^{hist}$ refers to the historical system state vector that carries statistical data and trend patterns with which the historical operation of a microgrid facility can be evaluated. As a result, analysis of system operational modes is possible. In addition, it can also be used as a basis for prediction of future service demands. This dynamic adjustment enables the conversion factor to reflect real-time system demand. When a service becomes scarce, its conversion factor automatically increases, thereby incentivizing more nodes to provide that service.

### *B. Verifiable Random Function*

The PoE mechanism employs Verifiable Random Functions (VRF) as the cryptographic method to achieve secure and fair block proposer selection. VRF, first introduced by Micali et al. [31], is a cryptographic approach that produces pseudorandom outputs with publicly verifiable proofs of correctness.

A VRF utilizes a private key and a specified input to generate a deterministic output and proof. This uniqueness prevents the proposer from "grinding" as can happen with simple pseudo-random number generators to select a preferred output. Each node can compute the VRF once per round, preventing them from selecting favorable outcomes. Anyone without the private key will see the VRF output as indistinguishable from uniform randomness. This property makes it impossible to predict future proposer so that targeted attacks and strategic manipulation are prevented. Until the seeds of these rounds are revealed, even the node itself cannot predict the future round's VRF output. The public key of any node can be used to verify that the VRF output was computed correctly without revealing the secret key. Because anyone on the network can verify the outcome, there is no need to rely on a trusted third party. This enables trustless protocol design.

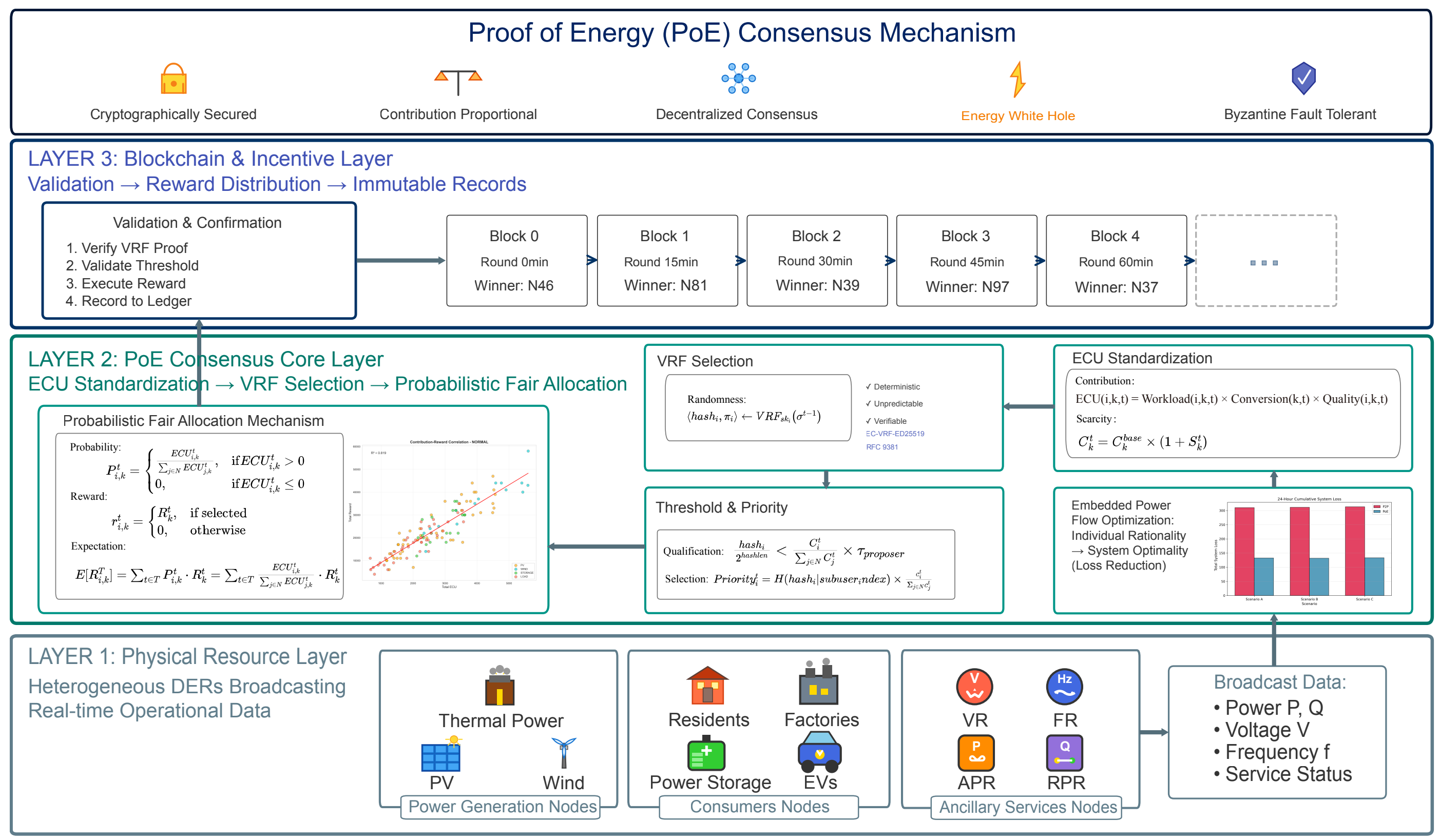

**Fig. 1.** The Architecture and Workflow of Proof of Energy

PoE employs the EC-VRF-ED25519 variant, which is based on Curve25519 elliptic curve cryptography and conforms to IETF RFC 9381 standard. On standard hardware, generating a VRF takes around 0.3-0.5 ms and verifying one takes 0.2-0.4 ms. The consensus cycle is set to 15 minutes to align with energy management requirements for energy systems. The security level for ECDLP on Curve25519 is 128 bits (i.e., equivalent to 3072-bit RSA security). This security level is appropriate for energy trading applications, where the cost of mounting an attack substantially exceeds any potential economic gain. RFC 9381 is an IETF standard that underwent extensive cryptographic review. The use of standardized protocols ensures interoperability and reduces implementation risk compared to custom cryptographic designs. The same EC-VRF-ED25519 implementation is deployed in Algorand's production blockchain, validating its real-world viability. [32]

The novelty of PoE lies not in using VRF for blockchain consensus as Algorand and others have done for a financial system but rather in adapting VRF to dynamic physical-world metrics. Traditional blockchain VRF applications utilize static weights that rarely change (e.g., PoS token holdings). In contrast, VRF selection weights in PoE are based on unit selection ECUs and are modified every fifteen minutes based on actual energy production, consumption, and provision of ancillary services. This adaptation presents several challenges including dynamic weight management, physical verification-based Sybil resistance, and convergence to fairness over time. Throughout Section III, we demonstrate that the equilibrium distribution of payoffs converges to a distribution proportional to contribution that is proportional to contribution ($R^2 > 0.75$ over 2880 rounds in the following experiments).

### C. VRF-based PoE Block Proposal Mechanism

The node's total contribution is the weighted sum of its normalized loads across all service dimensions.

$$C_i^t = \sum_{k=1}^{|S|} \alpha_k^t \times ECU_{i,k}^t \tag{4}$$

$\alpha_k^t$ represents the weight coefficient of service k at time t where $\sum_{k=1}^{|S|} \alpha_k^t = 1$. The weight coefficients indicate how important a service is to the system, which ultimately influences the comprehensive contribution. The service with a larger weight will have a greater impact here.

Service weights aren't fixed. They vary according to the scarcity conditions of the system.

$$\alpha_k^t = \frac{\alpha_k^{base} \times (1 + \beta \times S_k^t)}{\sum_{j=1}^{|S|} \alpha_j^{base} \times \left(1 + \beta \times S_j^t\right)} \tag{5}$$

where $\alpha_k^{base}$ denotes the baseline weight for service $k$, β represents the system response intensity parameter, and $S_k^t$

indicates the scarcity index of service $k$. This dynamic adjustment mechanism enables weights to reflect real-time system demands, directing resources toward more critically needed services.

The VRF-based proposer selection process consists of four phases: VRF generation, threshold validation, priority computation, and public verification. These phases ensure that selection is simultaneously contribution-proportional, unpredictable, and publicly verifiable.

Each node **i** independently computes its VRF output using its private key $sk_i$, which is securely generated during network initialization:

$$\langle hash_i, \pi_i \rangle \leftarrow VRF_{sk_i}(seed^{t-1}|\text{proposer}|round^t) \quad (6)$$

where, $hash_i \in \{0,1\}^{512}$: 64-byte (512-bit) pseudorandom output serving as the selection criterion. For anyone without $sk_i$, this value appears uniformly random, preventing prediction of future proposers. $\pi_i \in 0{,}1^{640}$: 80-byte (640-bit) cryptographic proof enabling any node to verify that $hash_i$ was correctly computed from the given input using node $i$'s public key $pk_i$, without revealing $sk_i$. $seed^{(t-1)}$: 32-byte random seed derived from the VRF output of the previous consensus round, ensuring that the current round's selection is unpredictable until the previous round completes. Proposer: Role identifier string distinguishing proposer selection from other potential VRF uses. The VRF function is deterministic---given identical input, it always produces the same $(hash_i, \pi_i)$ pair. However, without $sk_i$, the output is indistinguishable from random values. This deterministic yet unpredictable property is crucial: determinism enables verification (others can check correctness), while unpredictability prevents gaming (nodes cannot predict future selections to optimize behavior).

PoE implements VRF using the EC-VRF-ED25519 algorithm specified in RFC 9381. The function internally: (1) constructs input $M = seed^{(t-1)}|``\text{proposer}''|round^t$ ; (2) computes a signature-like value $\sigma = \text{ED25519}(sk_i, M)$; (3) derives $hash_i = \text{SHA}-512(\sigma)$; (4) formats proof $\pi_i$ from signature components. This construction inherits ED25519's security properties while achieving efficient verification.

Not all nodes become proposer candidates. Node i qualifies only when its VRF output satisfies the contribution-weighted threshold:

$$\frac{hash_i}{2^{hashlen}} < \frac{C_i^t}{\sum_{j \in N} C_j^t} \times \tau_{proposer} \quad (7)$$

This threshold condition implements contribution-proportional probabilistic selection. Division by $2^{hashlen}$ normalizes the VRF output to [0,1], enabling direct comparison with the contribution-based threshold. The parameter $\tau_{proposer}$ is typically set to 26, providing theoretical guarantee that at least one honest node will be selected with probability $1 - 10^{-11}$.

This design ensures: (1) contribution-proportionality - higher contributors have higher thresholds, thus higher qualification probability; (2) probabilistic selection - even low contributors have non-zero probability; (3) multiple candidates - $\tau = 26$ typically yields multiple qualifiers, necessitating a priority mechanism for final selection.

When several proposers come forth, the system chooses the block it adopts based on priority. Node $i$'s proposal priority is defined as:

$$Priority_i^t = H(hash_i|subuser_index) \times \frac{C_i^t}{\sum_{j \in N} C_j^t} \quad (8)$$

where $subuser_index$ handles multiple node selections and $H(\cdot)$ is a hash function. At the BA* stage, all the nodes run independently one after the other. Further, they choose proposal having highest value of $Priority_i^t$.

Chosen proposers need to include a proof $\pi_i$ of verifiable random function in their block proposals. Through this proof, other nodes can verify whether the proposal is legitimate by confirming the validity of the sortition proof and the correctness of priority calculations. This technique makes sure that everyone gets a fair chance.

The scheme announces the ultimate stability capacity and robust solvability with malicious contributors. Malicious proposers may send different block versions to different nodes. However, this only causes different nodes to start BA* with different initial values. Most likely, consensus will be achieved on an empty block. Assuming the honesty ratio of nodes satisfies $h > 2/3$, the probability that a Byzantine node is the proposer with the highest priority is at most $1 - h$. The adopted honest proposals will strengthen correctness as well.

After VRF generation, threshold validation, and priority computation are completed, all network nodes independently execute the public verification process to ensure the correctness and fairness of the consensus outcome. Verification comprises three key steps: 1) VRF Proof Correctness Verification: Each node uses the proposer's public key to verify VRF proof $\pi_i$ and confirms that $hash_i$ was indeed correctly computed from the corresponding private key $sk_i$ and input data. This verification process does not require knowledge of the private key, thereby achieving trustless public verifiability. 2) Threshold Compliance Verification: Nodes verify that the winning proposer's VRF output satisfies the threshold condition of equation (7), ensuring that the selection process strictly adheres to the contribution-proportionality principle. 3) Nodes check the selected block whether it has the maximum priority value (as per equation (8)), whenever there are qualified proposers. After the verification, the system will facilitate the probabilistic reward allocation.

In round **t** of the consensus, the implementation selects one receiver of the block reward uniformly at random using VRF. The chances of node **i** winning relies only on its total contribution value:

$$P_i^t = \begin{cases} \frac{C_i^t}{\sum_{j \in N} C_j^t}, & \text{if } C_i^t > 0 \\ 0, & \text{otherwise} \end{cases} \quad (9)$$

where $C_i^t$ is the total ECU contribution of node **i** in round **t** (computed by equation (4)), and **N** is the set of all active nodes.

The actual reward distribution adopts a "winner-takes-all" mechanism:

$$r_i^t = \begin{cases} R, & \text{if node } i \text{ is selected via VRF} \\ 0, & \text{otherwise} \end{cases} \quad (10)$$

where **R** is the block reward constant. This mechanism ensures atomic distribution of rewards per round, eliminates complex fractional payment mechanisms, and prevents targeted attacks and strategic gaming through VRF's unpredictability.

Although single-round reward allocation is stochastic, according to the law of large numbers, a node's expected cumulative reward over a sufficiently long time window **T** is strictly proportional to its cumulative contribution. The expected total reward for node **i** over time period **T** is:

$$E[R_i^T] = \sum_{t\in T} P_i^t \times R = \sum_{t\in T} \frac{c_i^t}{\sum_{j\in N} c_j^t} \times R \quad (11)$$

Further simplifying, if we define node **i**'s average contribution ratio as:

$$\bar{p}_i = \frac{1}{|T|}\sum_{t\in T} \frac{c_i^t}{\sum_{j\in N} c_j^t}$$

then:

$$E[R_i^T] = \bar{p}_i \times R \times |T| \quad (12)$$

where $|T|$ represents the total number of rounds in the time window.

This probabilistic mechanism has security benefits over the deterministic proportional allocation mechanism. It would be impossible for an attacker to foresee which node will obtain rewards in the future rounds. Since the property of a VRF, nodes won't be able to grind because they cannot compute the output with repeated computation. Each node will generate one deterministic VRF output for one round. The allocation of reward is fair, and any participant can check it by themselves not by any central authority or trusted third party.

This design exchanges unpredictability for security in the short term with a moderate degree of change; but long term, it achieves the fairness of contribution proportionality through mathematical expectation, realizing the incentive principle of " greater contribution greater return " while significantly enhancing the system security and attack resistance.

## III. Case Study

### A. Experimental Setup

Section III validates the proposed PoE consensus mechanism through simulation experiments. The experiments aim to: (1) verify the fairness of the ECU-based reward allocation mechanism; (2) evaluate the adaptability of dynamic conversion factors under different operating conditions; (3) analyze the incentive mechanism for flexible resources; (4) quantify the system efficiency improvement of PoE over conventional P2P trading; and (5) assess the convergence and robustness of the mechanism.

TABLE I
Experimental Parameters and Scenario Design

| Parameter | Normal | High | Purpose |
|---|---|---|---|
| Frequency mean | 50.00 Hz | 49.92 Hz | Frequency deviation effect |
| Frequency std. dev. | 0.02 Hz | 0.12 Hz | Volatility comparison |
| Renewable penetration | 40% | 70% | Penetration gradient |
| AGC trigger rate | ~5% | ~22% | Ancillary service demand |
| ECU weight (Energy: AS) | 90%:10% | 90%:10% | Service evaluation framework |
| Simulation duration | 30 days | 30 days | 2880 rounds/scenario |
| Monte Carlo runs | 10 | 10 | Statistical reliability |

The experiment network is based on the IEEE 123-node standard distribution system with modifications. A total of 108 active nodes is selected from the original IEEE 123-node feeder to represent a typical microgrid structure, while 15 passive relay nodes are excluded. The network operates at a rated voltage of 11 kV with a base power of 10 MVA and uses a three-zone radial topology. The system includes 48 PV nodes (0.5-2.0 MW), 15 wind power nodes (1.0-3.0 MW), 5 thermal generation nodes (2.0-5.0 MW), and 40 load nodes (0.5-1.5 MW). Two scenarios are designed: the Normal scenario represents standard operating conditions with 40% renewable penetration and stable frequency, while the High scenario represents stressed conditions with 70% renewable penetration and increased frequency deviation, as detailed in TABLE I.

The settlement interval is set to 15 minutes. Thermal nodes have a response time of 120 seconds with a speed compensation multiplier of 2.0, while PV and wind nodes have a response time of 30 seconds with a multiplier of 1.0. The block proposal process uses an Algorand EC-VRF-based cryptographic sortition mechanism with a proposal threshold $\tau_{proposer} = 26$. The comparison baseline is a P2P double auction mechanism, which matches trades based only on price without considering network topology constraints. System losses are calculated using the Newton-Raphson power flow algorithm (pandapower library), combining line active power losses, voltage deviation penalties (15%), and frequency deviation penalties (5%).

### B. Fairness Validation

The core design goal of the PoE mechanism is to achieve the fairness principle of reward proportional to contribution. Fig. 2 shows the overall performance comparison between the Normal and High scenarios, including fairness scatter plots, AS ratio time series, revenue heatmaps, and revenue structure comparison.

As shown in Fig. 2(a) and (b), the PoE mechanism achieves $R^2 = 0.9563$ in the Normal scenario and $R^2 = 0.9554$ in the High scenario, both well above the target threshold of 0.95. The data points are closely distributed along the ideal fairness line (red dashed line), which shows that node rewards are highly proportional to their ECU contributions. The $R^2$ difference between the two scenarios is only 0.0009, indicating that the fairness of the mechanism is not affected by operating conditions. The $R^2$ value means that 95.6% of the variance in node rewards can be explained by their ECU contributions, while the remaining 4.4% comes from the inherent randomness of VRF-based block selection.

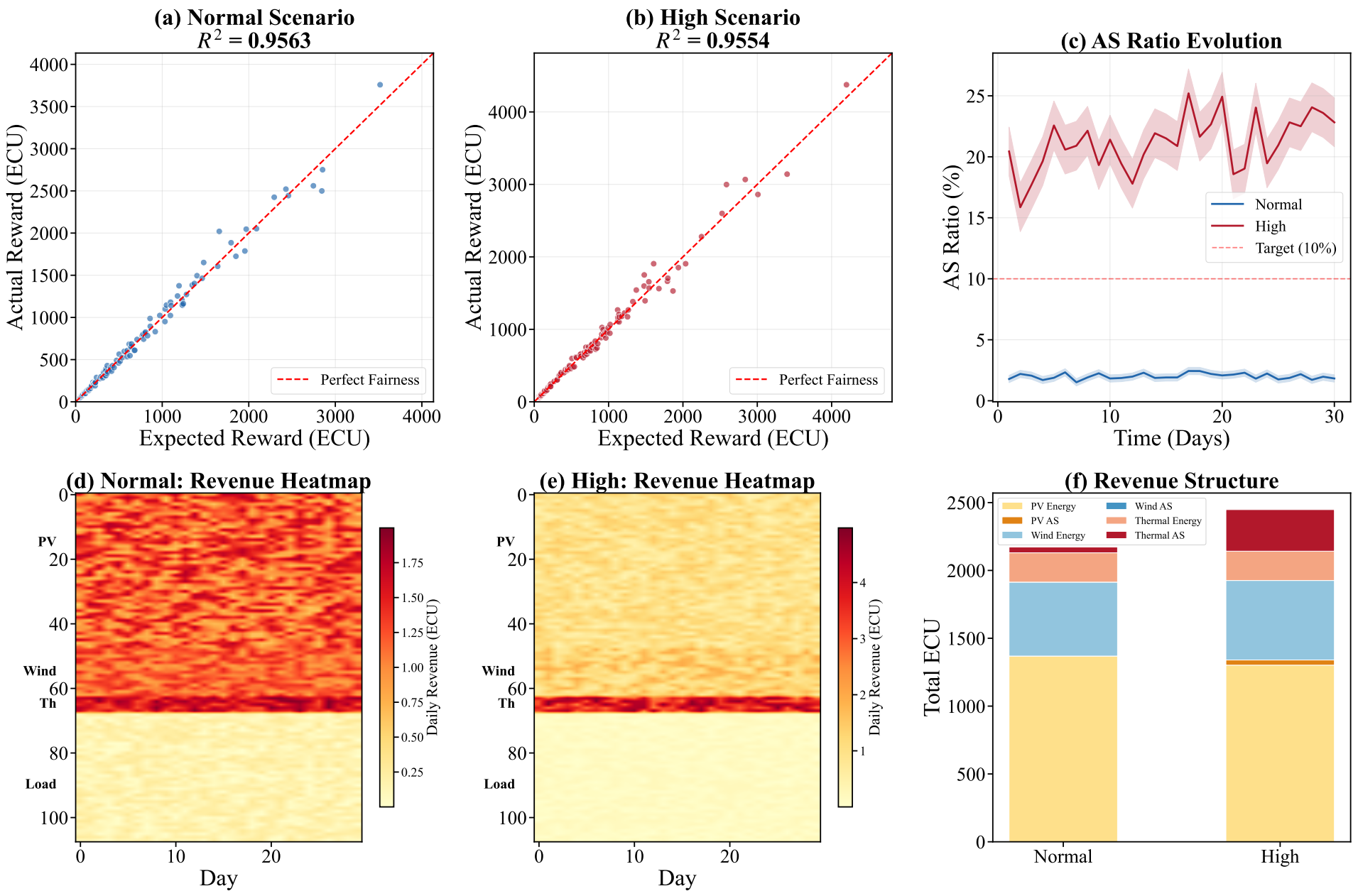

**Fig. 2.** Comprehensive performance comparison

The revenue heatmaps in Fig. 2(d) and (e) show the daily average revenue of each node over 30 days. In the Normal scenario, PV nodes (top 48 rows) display a stable orange-red band, reflecting steady energy service income. Thermal nodes (Th rows) show a dark red band due to additional ancillary service compensation. In the High scenario, the revenue of thermal nodes increases significantly, consistent with the higher demand for ancillary services. This spatial and temporal pattern confirms that the PoE mechanism maintains stable and consistent reward allocation over time.

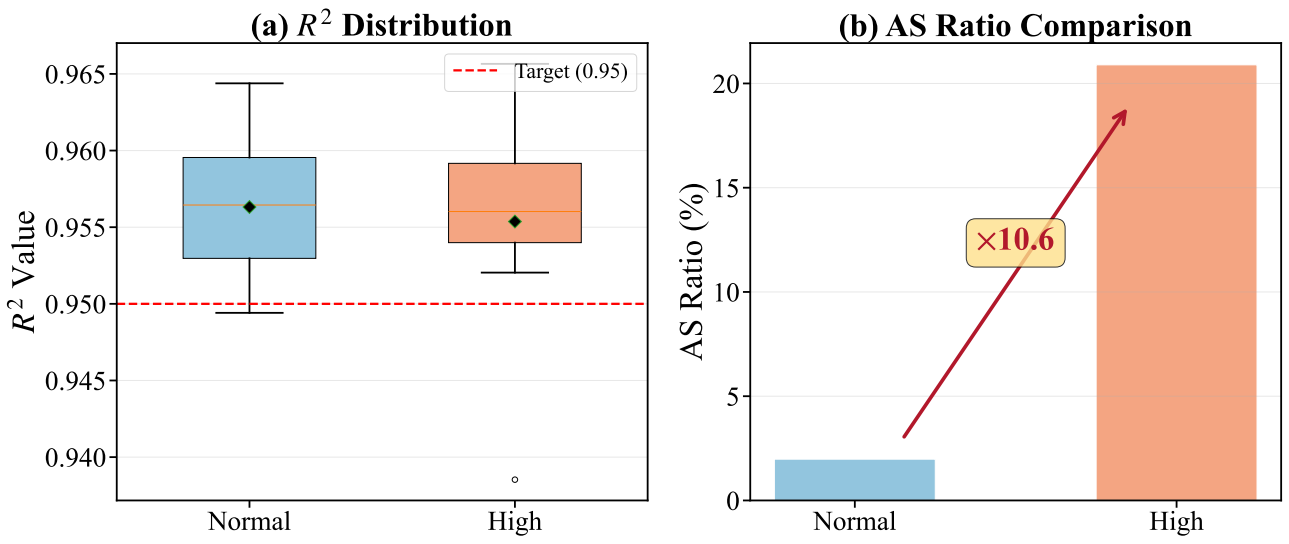

**Fig. 3.** Statistical validation. (a) R² distribution, (b) AS ratio distribution, (c) Statistical test summary.

Fig. 3 presents the statistical test results based on 10 Monte Carlo runs. The box plot in Fig. 3(a) shows that the $R^2$ values in both scenarios are above the 0.95 target line, with significant overlap between the two distributions. The t-test confirms no significant difference ($p = 0.736$). Fig. 3(b) is more important: the AS ratio increases from 2.0% in the Normal scenario to 20.9% in the High scenario, a 10.6-fold increase, showing a strong adaptive response of the dynamic conversion factors. This difference is highly statistically significant ($p < 0.001$). These results provide key evidence that the PoE mechanism is both stable in fairness and adaptive in pricing.

## *C. Adaptive Pricing Mechanism*

The dynamic conversion factor is a key innovation of the PoE mechanism. It automatically adjusts the ECU compensation for ancillary services based on the scarcity index. The AS ratio time series in Fig. 2(c) shows the dynamic behavior of this adaptive process. In the Normal scenario, the AS ratio stays at a low level of about 2%. In the High scenario, it fluctuates between 15% and 30%, reflecting real-time changes in system frequency deviation. This fluctuation pattern confirms that the dynamic conversion factor responds to system conditions in real time.

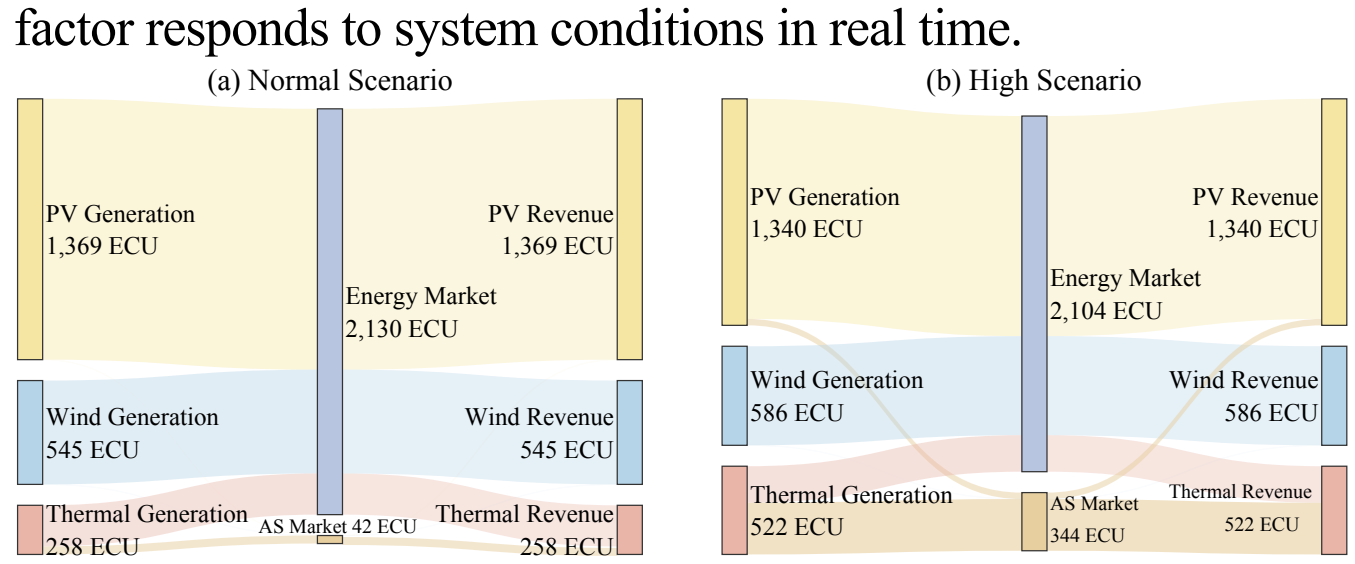

**Fig. 4.** Energy flow and revenue distribution under Normal/High scenario.

It is worth noting that this adaptive process is fully decentralized. Each node independently calculates its scarcity index based on locally observed frequency deviations and applies the same conversion factor formula defined by the protocol, producing consistent ECU measurements across the network. The Sankey diagrams in Fig. 4 show the structural difference in energy flow between the two scenarios. In the Normal scenario, the energy market is dominant, and the AS market channel is narrow. In the High scenario, the AS market channel expands significantly, with thermal nodes receiving 58.6% of their revenue through the AS market. This forms a dual-channel revenue structure combining energy and ancillary services. The adaptive shift in revenue structure validates the effectiveness of the dynamic conversion factor as a decentralized price signal.

## *D. Flexible Resource Incentive*

Under high renewable energy penetration, flexible resources such as thermal generators are essential for maintaining system stability. However, their energy market revenue is reduced by the low marginal cost of renewables (merit-order effect). The PoE mechanism provides additional incentives for flexible resources through ancillary service ECU compensation and response speed compensation multipliers.

TABLE II
THERMAL NODE REVENUE DECOMPOSITION (PER NODE)

| Scenario | Energy ECU | AS ECU | Total ECU | AS Share |
|---|---|---|---|---|
| Normal | 43.2 (83.7%) | 8.4 (16.3%) | **51.6** | 16.3% |
| High | 43.2 (41.4%) | 61.2 (58.6%) | **104.4** | 58.6% |
| Variation | 0% | **+629%** | **+102%** | +42.3 pp |

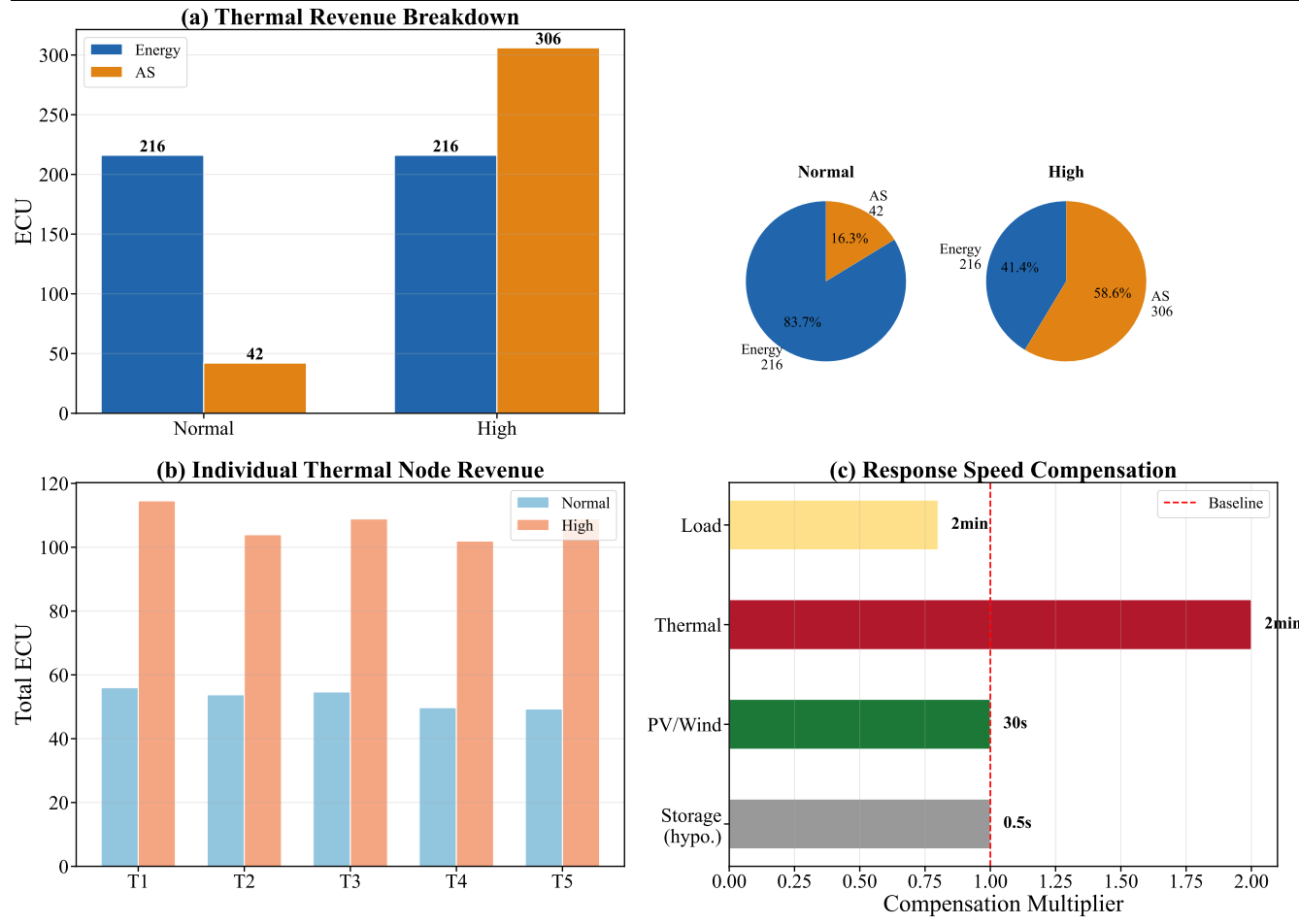

**Fig. 5.** Thermal node performance analysis.

As shown in TABLE II and Fig. 5, the energy ECU of thermal nodes remains the same in both scenarios (43.2 ECU), but the AS ECU increases from 8.4 to 61.2, doubling the total ECU. The pie charts in Fig. 5 clearly show this structural shift: thermal revenue changes from energy-dominated (83.7%) to AS-dominated (58.6%). Fig. 5(c) shows the response speed compensation mechanism. The 120-second response time of thermal nodes corresponds to a compensation multiplier of 2.0, which is much higher than the 1.0 baseline for PV and wind nodes. This reflects the irreplaceable value of slow-response resources in providing sustained frequency regulation services.

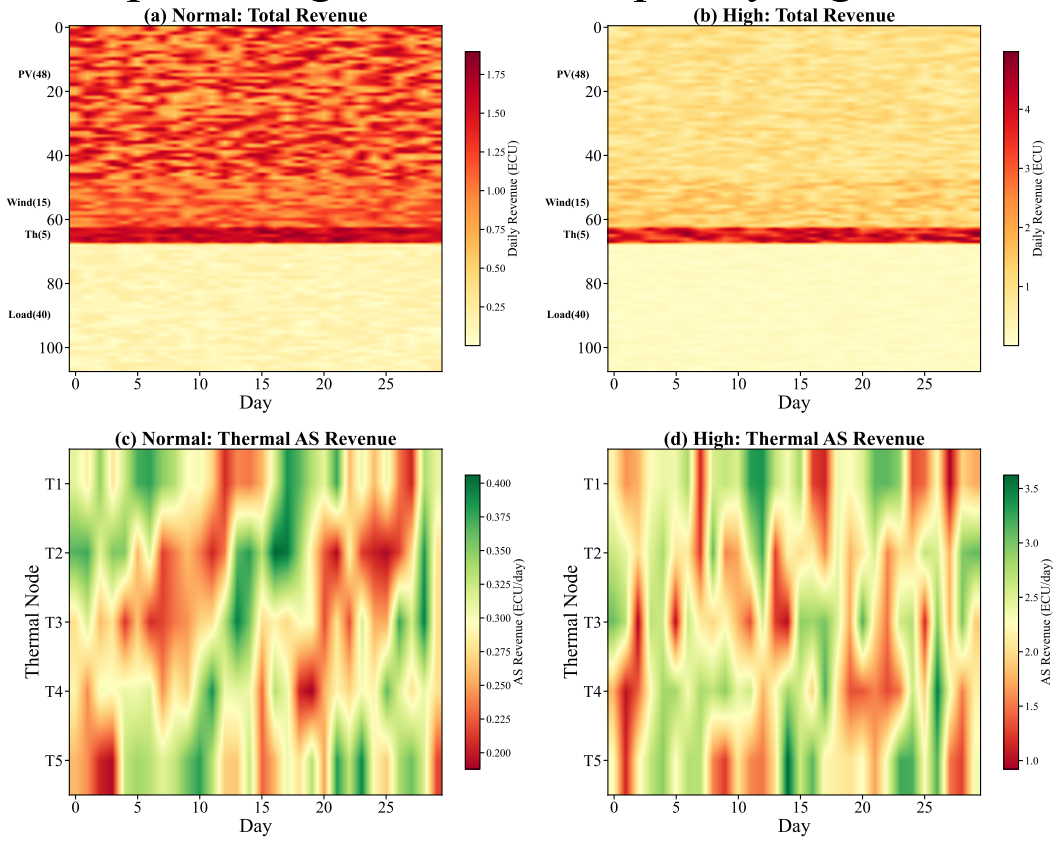

**Fig. 6.** Revenue evolution heatmaps.

Fig. 6 provides a more detailed view of the spatial and temporal distribution. The thermal AS revenue heatmaps in Fig. 6(c) and (d) are particularly noteworthy. In the Normal scenario, AS revenue ranges from 0.24 to 0.34 ECU/day with small variations. In the High scenario, the range expands to 1.25-3.0 ECU/day with clear temporal patterns, reflecting intra-day and inter-day variations in system frequency deviation. These patterns are a direct result of the PoE dynamic pricing mechanism responding to actual system fluctuations, rather than pre-set fixed parameters.

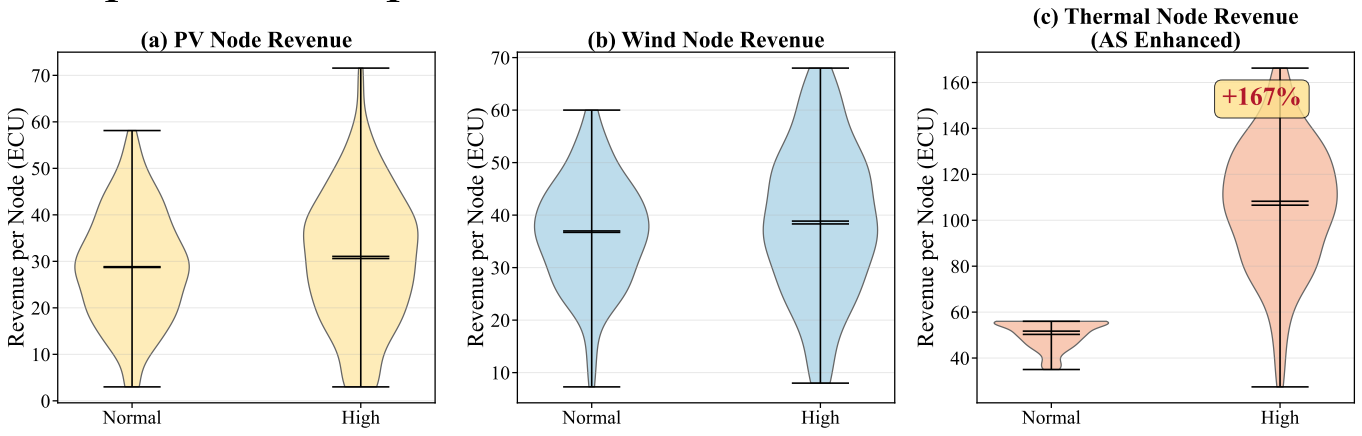

**Fig. 7.** Node revenue distribution (violin plots). (a) PV nodes, (b) Wind nodes, (c) Thermal nodes revenue enhancement under High scenario with significantly expanded distribution.

The violin plots in Fig. 7 show the statistical distribution of revenue for each node type. PV nodes (Fig. 7(a)) have similar distributions in both scenarios with a median of about 25 ECU, reflecting the stability of energy service revenue. Wind nodes (Fig. 7(b)) show a much wider distribution in the High scenario, reflecting the impact of wind power output fluctuations on revenue under high penetration. The most notable result is for thermal nodes (Fig. 7(c)): the distribution in the High scenario expands significantly, from a narrow range of 0-100 ECU to a wide range of 0-340 ECU, an increase of +167%. This clearly demonstrates the enhanced incentive effect of the PoE mechanism on flexible resources.

## *E. System Loss Comparison*

To evaluate the system efficiency of the PoE mechanism, a loss comparison experiment is conducted against a P2P double auction mechanism. The P2P auction uses a price-priority matching rule, considering only buyer and seller bids without incorporating network topology information. Both mechanisms use the same 108-node network, node configuration, and time series data.

TABLE III
SYSTEM LOSS COMPARISON: POE VS. P2P AUCTION

| PoE (kW) | P2P (kW) | Improvement | p-value |
|---|---|---|---|
| 2382 ± 198 | 2747 ± 236 | **13.3%** | < 0.001 |

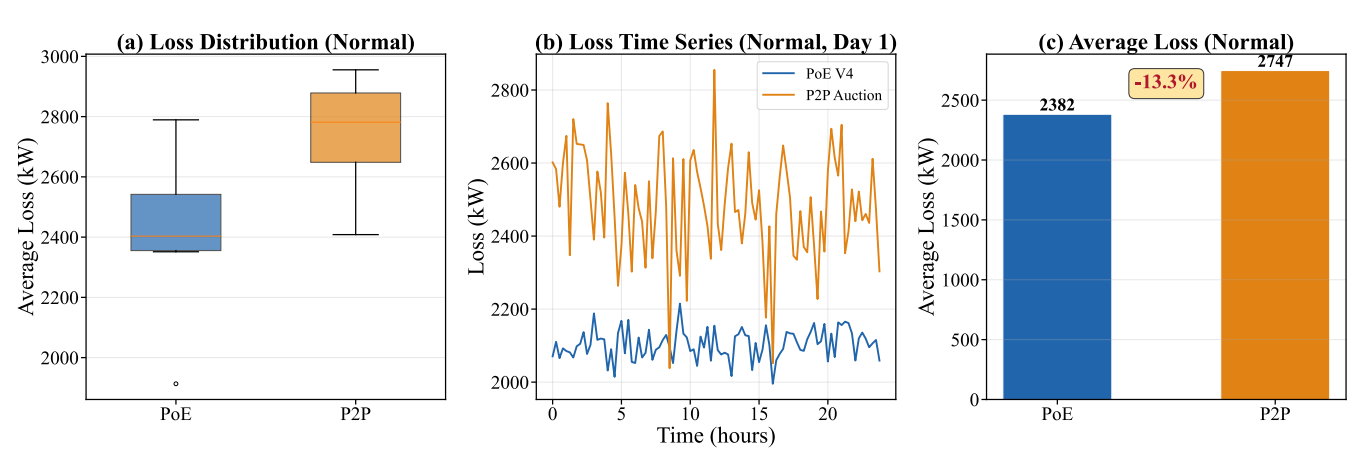

**Fig. 8.** System loss comparison: PoE vs P2P auction. (a) Average loss distribution (box plot), (b) Loss time series comparison, (c)Average Loss (Normal)

As shown in TABLE III and Fig. 8, the PoE mechanism achieves an average loss of 2,382 kW in the Normal scenario, which is 13.3% (365 kW) lower than the 2,747 kW of the P2P auction. The statistical test confirms that this difference is highly significant ($p < 0.001$). The loss reduction is mainly due to the incentive effect of ECU: nodes tend to optimize their local power injection to maximize ECU revenue, which results in a more balanced power distribution across the network. In contrast, the price-only matching of the P2P auction ignores network topology constraints, which can cause power flows to concentrate on long-distance lines.

The time series comparison in Fig. 8(b) further reveals the difference between the two mechanisms. The loss curve of the P2P auction fluctuates significantly (standard deviation 120-185 kW), reflecting the unstable power distribution caused by random price matching. In contrast, the PoE loss curve is smooth (standard deviation 34-48 kW), showing the continuous optimization effect of ECU incentives on power distribution. It is also noted that the loss performance of PoE is nearly identical across both scenarios (difference < 0.01%), indicating that the loss reduction mainly comes from improved power allocation patterns rather than frequency response. This gives PoE a robustness advantage across different operating scenarios.

### *F. VRF Consensus Performance*

The Algorand EC-VRF-based cryptographic sortition mechanism is a core component of the decentralized security of PoE. To independently verify its performance, the VRF selection process over 2,880 rounds (108 nodes, 311,042 generation operations in total) is analyzed.

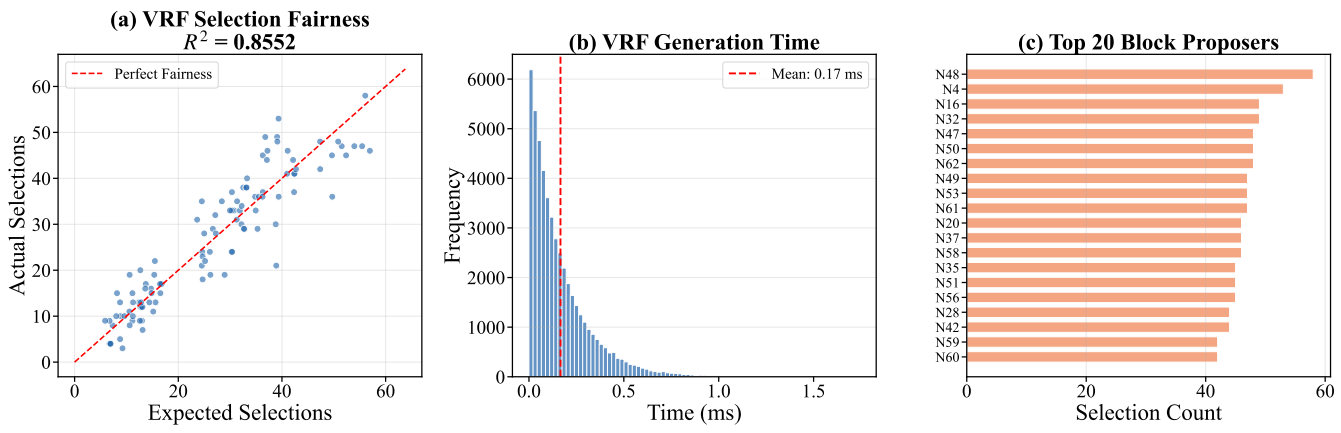

**Fig. 9.** VRF performance and fairness analysis. (a) Selection fairness: expected vs actual selections, (b) VRF generation time distribution, (c) Top 20 block proposers showing balanced selection.

Fig. 9(a) shows that the VRF selection fairness $R^2 = 0.8552$. This value is lower than the reward allocation $R^2$ (0.956) because they measure fairness at different levels: VRF selection is a single-round random event with naturally higher variance, while reward allocation is a 30-day cumulative result that achieves high proportionality through the law of large numbers. Fig. 9(b) shows that the VRF generation time is concentrated around 0.17 ms, well below the 15-minute settlement interval, ensuring the feasibility of real-time trading. The top 20 block proposer distribution in Fig. 9(c) shows that the selection count for each node falls within the range of 42-58, with no significant monopoly, confirming the decentralized nature of VRF. The verification success rate reaches 99.999%, ensuring cryptographic security.

### *G. Convergence and Robustness*

The practical deployment of the PoE mechanism requires that its fairness metrics converge to a stable level within a limited time.

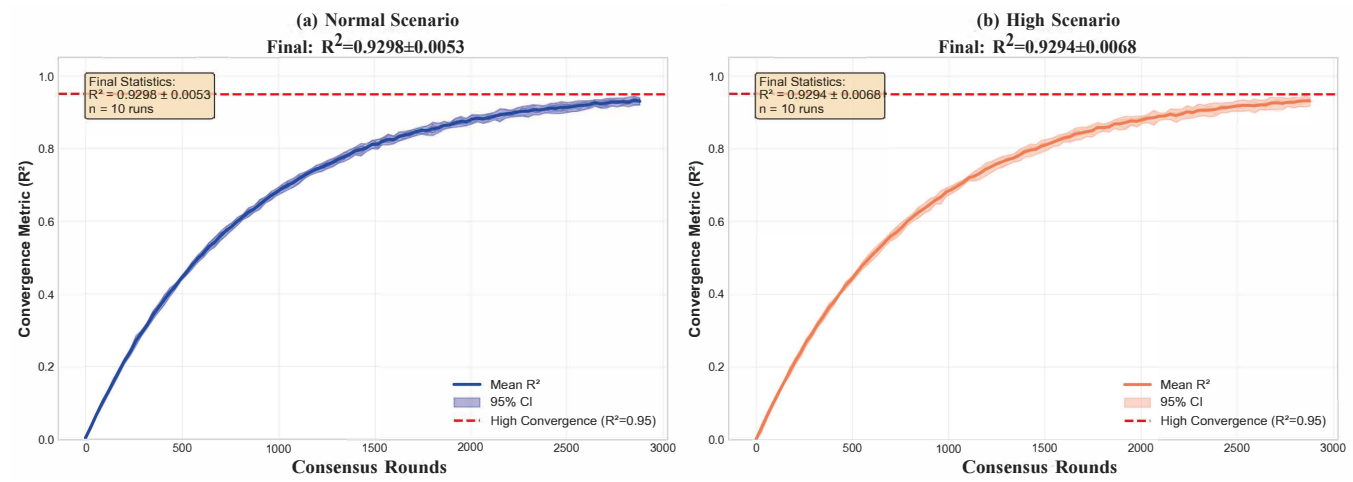

**Fig. 10.** $R^2$ convergence trajectories. (a) Normal scenario, (b) High scenario.

The $R^2$ values in both scenarios show an exponential convergence pattern, reaching the 0.90 threshold at about 800 rounds (approximately 8.3 days) and then gradually stabilizing. The Normal scenario converges to a final $R^2 = 0.9298 \pm 0.0053$, and the High scenario converges to $R^2 = 0.9294 \pm 0.0068$. The convergence paths of the two scenarios nearly overlap completely, and the t-test confirms no significant difference in the final $R^2$ values ($t = 0.343$, $p = 0.736$). It should be noted that the standard deviation of the AS ratio in the Normal scenario is close to zero ($\sigma \approx 0.000$). This is because the frequency deviation in this scenario is very small ($\sigma = 0.02$ Hz), making the AGC trigger pattern highly deterministic. This further confirms the deterministic response characteristic of the PoE pricing mechanism.

TABLE IV
COMPREHENSIVE STATISTICAL SUMMARY

| Metric | Normal | High | Statistical Test |
|---|---|---|---|
| Fairness $R^2$ | 0.9563 | 0.9554 | p = 0.736 (n.s.) |
| AS Ratio | 2.0% | 20.9% | p < 0.001 *** |
| Convergence $R^2$ (30d) | 0.9298 ± 0.005 | 0.9294 ± 0.007 | p = 0.736 (n.s.) |
| VRF Selection $R^2$ | 0.8552 | — | 311,042 ops |
| VRF Gen. Time | 0.17 ms | — | P95: 0.17 ms |
| VRF Verification Rate | 99.999% | — | 311,043 verif. |
| Loss Improvement (vs P2P) | 13.3% | — | p < 0.001 |
| PoE Loss Std. Dev. | 34-48 kW | — | Vs P2P: 120-185 kW |

## IV. CONCLUSION

This paper proposes a Proof of Energy (PoE) blockchain consensus mechanism that addresses the four fundamental challenges in decentralized microgrid electricity trading. For the coordination problem, PoE achieves decentralized coordination through scarcity indices embedded in the protocol rules. Each node independently calculates local frequency deviations and produces consistent ECU pricing. The AS ratio adapts by a factor of 10.6 across scenarios, confirming the ability of "local sensing, global consistency" coordination without a central dispatcher. For the economic problem, the ECU-based unified service evaluation framework converts heterogeneous services into a comparable metric. The reward allocation fairness reaches $R^2 = 0.956$, and thermal AS revenue increases by 629%, resolving the long-standing issue of ancillary service undervaluation in conventional mechanisms. For the trust problem, the EC-VRF-ED25519-based cryptographic sortition binds block proposal rights to verifiable energy contributions, achieving a verification success rate of 99.999% and eliminating the possibility of data falsification and identity manipulation. For the optimization problem, the ECU incentive creates an incentive-compatible mechanism where self-interested node behavior results in a more balanced power distribution, reducing system losses by 13.3% compared to P2P auction ($p < 0.001$). However, it should be noted that this improvement is an indirect result of incentive guidance rather than explicit optimization, leaving room for further efficiency gains. Future work will focus on three directions: prototype deployment in real microgrid environments, game-theoretic analysis of mechanism robustness under malicious behavior, and scalability studies in larger distribution networks.